\documentclass[aps,preprint,showpacs,superscriptaddress,groupedaddress]{revtex4}  
\usepackage{graphicx}  
\usepackage{dcolumn}   
\usepackage{bm}        
\usepackage{amssymb}   
\usepackage{amsmath}   

\begin{document}


\title{Low-scale seesaw models versus $N_{\rm eff}$}

\date{\today}

\author{P.~Hern\'andez}
\author{M.~Kekic}
\affiliation{IFIC (CSIC-UVEG), Edificio Institutos Investigaci\'on, 
Apt.\ 22085, E-46071 Valencia, Spain}
\author{J. L\'opez-Pav\'on}
\affiliation{SISSA, via Bonomea 265, 34136 Trieste, Italy}
\affiliation{ INFN, Sezione di Trieste, 34126 Trieste, Italy.}

\preprint{IFIC/13-81}
\preprint{SISSA  50/2013/FISI}

\begin{abstract}
We consider the contribution of the extra sterile states in generic low-scale seesaw models to 
extra radiation, parametrized by $N_{\rm eff}$. We find that the value of $N_{\rm eff}$ is  roughly independent 
of the seesaw scale within a wide range. We explore the full parameter space in the 
case of two extra sterile states and find that these models  are strongly constrained by cosmological data
for any value of the seesaw scale below ${\mathcal O}(100$MeV). 
\end{abstract}
\maketitle


Models with extra light sterile neutrinos with masses in the range of ${\mathcal O}(1$eV) 
could provide an explanation to some of the 
neutrino anomalies \cite{Kopp:2013vaa}, such as the appearance signal $\bar{\nu}_\mu \rightarrow \bar{\nu}_e$ of the LSND experiment \cite{Aguilar:2001ty}, undisproved by the MiniBOONE \cite{Aguilar-Arevalo:2013pmq} experiment, or the deficit 
of neutrinos ($\bar{\nu}_e \rightarrow \bar{\nu}_e$) in short-baseline reactor experiments, the so-called reactor neutrino anomaly \cite{reactor}. Sterile species in the keV range
could still be valid candidates for warm dark matter \cite{Dodelson:1993je}, while species in the GeV range could account for the baryon asymmetry in the Universe \cite{Akhmedov:1998qx}.

Models with $N$ extra sterile states are usually defined as phenomenological models with a generic neutrino mass matrix of size $3+N$ without 
specifying whether neutrinos are Dirac or Majorana. In the former case, a renormalizable Lagrangian representing this model
 would require the addition of $3+2 N$ extra singlet Weyl fermions to the minimal Standard Model (SM) so that they can be
 paired up into $3+N$ Dirac neutrinos.  In contrast, if neutrinos 
 are Majorana, such model is necessarily an effective low-energy theory.  We can insist on having $3+N$ Majorana fermions only 
 and  a renormalizable Lagrangian, but in this case the mass matrix will not be generic, since Majorana entries for the charged neutrinos are forbidden
 by the gauge symmetry. We have in this case the so-called mini-seesaw \cite{miniseesaw} or minimal models \cite{Donini:2011jh}. These are simply the standard Type I seesaw models with   a low (ie. below electroweak) Majorana mass scale. The generic feature of these
 models is that active-sterile mixings are strongly correlated with the ratio of the light-to-heavy masses. They are therefore much more
 constrained (ie. they have  less free parameters than the phenomenological models). It should be stressed that seesaw models are the simplest extensions of the SM to accommodate massive neutrinos, but they can 
 do so independently of the value of the seesaw scale (ie. the scale of Majorana masses).
  
 It has been pointed out that the neutrino anomalies  could also be accounted for in these minimal models if $N \geq 2$ (with the same caveats as in the phenomenological models) in spite of the strong
correlation between mixings and mass splittings \cite{Donini:2012tt}. In other words, the order of magnitude for the active-sterile mixing given by the seesaw limit for a seesaw scale
of ${\mathcal O}(1$eV) is in the right ballpark to explain the neutrino anomalies, which is remarkable. These minimal models with $N=3$, and a much higher seesaw scale, have also been proposed as candidates
to explain dark matter and the baryon asymmetry \cite{Canetti:2012kh}. 

It is well-known that light sterile neutrinos with significant active-sterile mixing can be strongly constrained by cosmological 
measurements. The energy density of the extra neutrino species, $\epsilon_s$, is usually quantified in terms of $N_{\rm eff}$ (when they are relativistic) defined by 
\begin{equation}
 N_{\rm eff} \equiv {\epsilon_{s} + \epsilon_\nu \over \epsilon^0_\nu},
\end{equation}
where $\epsilon^0_\nu$ 
is the energy density of one SM massless neutrino with a thermal distribution (below $e^\pm$ annihilation it is $\epsilon^0_\nu \equiv (7 \pi^2/120) (4/11)^{4/3} T_\gamma^4$  at the photon temperature $T_\gamma$). In the minimal SM with massless neutrinos $N_{\rm eff} = 3.046$ at CMB \cite{Mangano:2005cc}. One fully thermal extra sterile state that decouples being relativistic contributes $\Delta N_{\rm eff} \simeq 1$ when it decouples. 

$N_{\rm eff}$ at big bang nucleosynthesis (BBN) strongly influences
the  primordial helium production. A recent analysis of BBN bounds \cite{Izotov:2010ca} gives $N_{\rm eff} ^{BBN} = 3.68(3.80)^{0.80}_{-0.70}$ at $2 \sigma$, where the central value depends
 on the choice for the neutron lifetime, and assumes no lepton asymmetry. 
$N_{\rm eff}$ also affects the anisotropies of  the cosmic microwave background  (CMB).   
Recent CMB measurements
from Planck give  $N_{\rm eff}^{\rm CMB} = 3.30 \pm 0.27 (1\sigma)$ \cite{Ade:2013zuv}, which includes WMAP-9 polarisation data \cite{WMAP} and high multipole measurements from the South Pole Telescope 
\cite{SPT} and the Atacama Cosmology Telescope \cite{ACT}.

The contribution of extra sterile states to $N_{\rm eff}$ within  phenomenological models  has been extensively studied \cite{Dolgov:2003sg}-\cite{Melchiorri:2008gq}. For recent analyses see \cite{Hannestad:2012ky}-\cite{Archidiacono:2013xxa}. In particular the models that could accommodate the neutrino anomalies seem to be in strong tension with cosmology, specially those with two extra species. 

The purpose of this paper is to evaluate $N_{\rm eff}$ in the context of the much more constrained minimal seesaw models. Interestingly in spite of the fact that
the active-sterile mixings decrease with increasing seesaw scale, the rate of thermalisation of the sterile neutrinos is roughly independent of that scale. The bounds therefore apply in a wide range of seesaw scales.

 \paragraph{Thermalization in minimal $3+N$ models.}

   The minimal models are described by the most general renormalizable Lagrangian including $N$ extra singlet Weyl fermions, $\nu_R^i$:
   \begin{eqnarray}
{\cal L} = {\cal L}_{SM}- \sum_{\alpha,i} \bar L^\alpha Y^{\alpha i} \tilde\Phi \nu^i_R - \sum_{i,j=1}^N {1\over 2} \bar\nu^{ic}_R M_N^{ij} \nu_R^j+ h.c., \nonumber
\label{eq:lag}
\end{eqnarray}
where $Y$ is a $3\times N$ complex matrix and $M_N$ a diagonal real matrix. The model with $N=1$, that contains only two massive states, cannot explain the measured neutrino masses and mixings \cite{Donini:2011jh}. 
For $N=2$, the spectrum contains four massive states and one massless mode, whose mixing is described by four angles and three physical CP phases. For $N=3$, there are six massive states and the mixing is described in terms of six angles and six CP phases.  We will concentrate on the simplest model that can explain neutrino  data, i.e. $N=2$. The case with $N=3$ will be 
considered elsewhere.

We assume that the eigenvalues of $M_N$ are significantly larger than the atmospheric and solar neutrino mass splittings, which implies a hierarchy $M_N\gg Y v$ and therefore the seesaw approximation is good.  A convenient parametrization in this case is provided by that of Casas-Ibarra \cite{Casas:2001sr}, or its extension to all orders in the seesaw expansion as described in \cite{Donini:2012tt} (for an alternative see \cite{Blennow:2011vn}). 
The mass matrix can be written as
\begin{eqnarray}
{\mathcal M}_\nu = U^*~ Diag(m_l,M_h)~ U^\dagger.
\end{eqnarray}
where $m_l$ is a diagonal matrix with a zero and the two lighter masses, and $M_h$ contains the $N$ heaviest. Denoting by $a$ the active/light neutrinos and $s$ the sterile/heavy species, the  unitary matrix can be written as
\begin{eqnarray}
U =  \left(\begin{array}{lll}  U_{aa} & U_{as} \\
U_{sa} & U_{ss} 
\end{array}\right), 
\label{eq:u5}
\end{eqnarray}
with
\begin{eqnarray}
U_{aa} &=& U_{PMNS} \left(\begin{array}{ll} 1 & 0\\
0& H \end{array}\right), ~U_{ss} = \overline{H}, \nonumber\\
 U_{sa} &=&  i \left( \begin{array}{ll} 0 & \overline{H} M_h^{-1/2} R m_l^{1/2}  
 \end{array}\right), \nonumber\\
 U_{as} &=& i U_{PMNS} \left( \begin{array}{l} 0\\
  H m_l^{1/2} R^\dagger M_h^{-1/2} \end{array}\right),
  \label{eq:param}
\end{eqnarray}
where $U_{PMNS}$ is a $3\times 3$ unitary matrix, $R$ is a generic $2 \times 2$ orthogonal complex  matrix, while $H$ and $\bar{H}$ are 
defined by
\begin{eqnarray}
H^{-2} = I + m_l^{1/2} R^\dagger M_h^{-1} R m_l^{1/2}, \;\;\; \nonumber\\
 \overline{H}^{-2} = I + M_h^{-1/2} R m_l R^\dagger M_h^{-1/2}.
 \label{eq:hbar}
\end{eqnarray}
At leading order in the seesaw expansion, i.e. up to ${\mathcal O}\left({m_l\over M_h}\right)$, $H\simeq \overline{H} \simeq 1$, and we
recover the Casas-Ibarra parametrization.

The measured neutrino masses and mixings fix most of the parameters in these models. The only free parameters are two CP phases
of $U_{PMNS}$ that are presently unconstrained, the matrix $R$ that depends on a complex angle and the two heavy masses in $M_h$. 

The active neutrinos in the minimal SM are in thermal equilibrium in the early universe at temperatures above ${\mathcal O}(1$MeV). The presence
of extra singlets can modify the value of $N_{\rm eff}$ because the active-sterile mixing can also bring the singlets into thermal equilibrium. Obviously the thermalisation process depends very strongly on the mixing parameters and the neutrino masses. We assume throughout that neutrinos are relativistic. 

In \cite{simple:1990} a simple estimate for the thermalisation of one sterile neutrino was given as follows. Assuming that the active neutrinos are in thermal equilibrium with a collision rate 
given by $\Gamma_a$, the collision rate for the sterile neutrinos can be estimated to be
\begin{eqnarray}
\Gamma_{s_i} \simeq  {1 \over 2} \sum_ a \langle P(\nu_a \rightarrow \nu_{s_i})\rangle \times \Gamma_a,
\label{eq:gammas}
\end{eqnarray}
where $\langle P(\nu_a \rightarrow \nu_s)\rangle$ is the time-averaged  probability 
$\nu_a \rightarrow \nu_s$ (the factor $1/2$ results from a  more detailed analysis, see below). This probability depends strongly on temperature
because the neutrino index of refraction in the early universe is modified by coherent scattering of neutrinos with the particles in the plasma \cite{Notzold:1987ik}.
Thermalization will be achieved if there is any temperature where this rate is higher than the Hubble expansion rate $\Gamma_s(T) \geq H(T)$.  One can therefore
find the maximum of the function $f_s(T)\equiv \Gamma_s(T)/H(T)$ as function of $T$ and estimate $N_{\rm eff} \simeq N_{\rm eff}^{SM}+\sum_i \left(1-\exp(-\alpha f_{s_i}(T^i_{max}))\right)$ at decoupling, where $\alpha$ is an ${\mathcal O}(1)$ numerical constant. 
The Hubble expansion rate is $H(T) = \sqrt{{4 \pi^3 g_*(T) \over 45}} {T^2\over M_{\rm Planck}}$, where $g_*(T)$ is a function of the temperature.

Employing the method described in \cite{KTY} we find the time-averaged  probabilities in the primeval plasma  to be approximately
\begin{eqnarray}
 \langle P(\nu_a \rightarrow \nu_{{s}_i}) \rangle &=& 2 \left(\frac{M^2_{i}}{2 p V_a-M^2_{i}}\right)^2|U_{a s_i}|^2
 + \mathcal{O}\left(U^4_{as} \right),\nonumber\\
\end{eqnarray} 
where $p$ is the neutrino momentum and $V_a \equiv  A_a T^4 p$, with $A_e=A$, while $A_{\mu/\tau}= B$ for $T$ below the $\mu/\tau$ threshold ($T\lesssim 20/180$ MeV) or $A_{\mu/\tau}= A$ for higher $T \gtrsim 20/180$ MeV, where 
\begin{eqnarray}
B &\equiv&-2\sqrt{2}\,\left(\frac{7\zeta(4)}{\pi^2}\right)\frac{G_F}{M_Z^2},\nonumber\\
A&\equiv&B -4\sqrt{2}\,  \left(\frac{7\zeta(4)}{\pi^2}\right)\frac{G_F}{M_W^2}.
\end{eqnarray}

A more detailed description is provided by the density matrix formalism \cite{Sigl:1992fn,Dolgov:2002wy}:
\begin{eqnarray}
\dot\rho = - i [ \hat{H},\rho] - {1 \over 2} \{\Gamma, \rho-\rho_{eq} I_{A}\},
\label{eq:boltz}
\end{eqnarray}
where $\hat{H}$ is the Hamiltonian describing the propagation of relativistic neutrinos in the plasma, which in the flavour basis is given by
\begin{eqnarray}
{\hat H} = U^* {\rm Diag}\left({m^2_l\over 2 p}, {M^2 _h\over 2 p}\right)U^T + {\rm Diag}(V_e, V_\mu, V_\tau, 0,0),
\end{eqnarray}
and the collision term $\Gamma = {\rm Diag}(\Gamma_e,Ê\Gamma_\mu, \Gamma_\tau,0,0)$ 
\begin{eqnarray}
\Gamma_a= y_a {180 \zeta(3)\over 7 \pi^4} G_F^2 T^4 p, 
\end{eqnarray}
with $y_e= 3.6$, and $y_\mu = y_\tau= 2.5$ below the corresponding $\mu$ and $\tau$ thresholds, becoming equal to $y_e$ above \cite{ys}.  Finally $\rho_{eq}$ is the Fermi-Dirac distribution and $I_A = {\rm Diag}(1,1,1,0,0)$.

Separating the equations into the active $A$  and sterile $S$ blocks and assuming that 
$\Gamma_a(T) \gg H(T)$, collisions are then fast enough to equilibrate $\rho_{AA}$ and $\rho_{AS}$, ie. $\dot\rho_{AA}=\dot\rho_{AS}=0$ (the so-called 
static approximation \cite{Dolgov:2003sg}). If we assume hierarchical heavy masses, and take into account the seesaw expansion, it is possible to show  that the thermalisation of the different sterile states
approximately decouple, and the equation for each species simplifies to
\begin{eqnarray}
\dot{\rho}_{ss} &=& 
-   \left(H_{AS}^\dagger \left\{{\Gamma_{AA} \over  (H_{AA}- H_{ss} )^2 +   \Gamma^2_{AA}/4}  \right\}H_{AS}\right)_{ss} \tilde\rho_{ss}\nonumber\\
&\simeq& - {1 \over 2} \sum_a  \langle P(\nu_s \rightarrow \nu_a) \rangle \Gamma_a \tilde{\rho}_{ss}, 
\label{eq:dec}
\end{eqnarray}
where $\tilde{\rho}_{ss} \equiv \rho_{ss}- \rho_{eq}$.  This equation justifies the estimate of eq.~(\ref{eq:gammas}).

$T_{max}$  is the value of the temperature at which $\Gamma_s(T)/H(T)$ is maximum. Taking $p \simeq 3.15 T$, it is easy to see that for each sterile state of mass $M_i$, $T_{\max}$ can be bounded by 
\begin{eqnarray}
\left({M_i^2\over 59.5~ |A_e|}\right)^{1/6} \leq T_{max} \leq \left({M_i^2\over 59.5 ~Ê|A_\tau|}\right)^{1/6},
\end{eqnarray}
so it depends significantly on $M_i$ but weakly on the mixings. Taking into account the seesaw scaling $|U_{a s_i}|^2 \sim {\mathcal O}(m_l/M_i)$, it follows that $f_{s_i}(T_{\rm max})$ 
is roughly independent of $M_i$.

\paragraph{$N_{\rm eff}$ in minimal $3+2$ models.}

\begin{figure}
\includegraphics[scale=1]{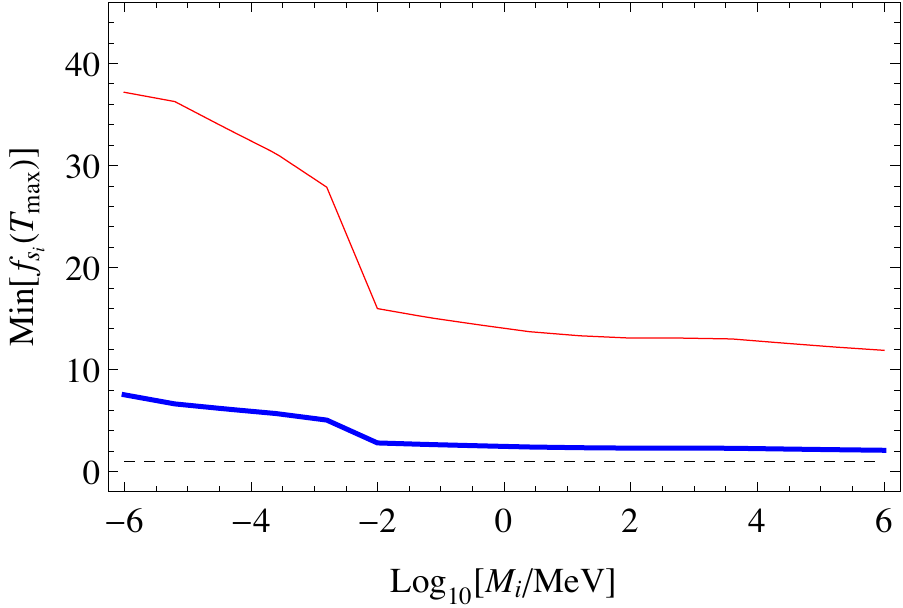}
\caption{\label{fig:fmin} ${\rm Min}[f_{s_i}(T_{\rm max})]$ for the lighter sterile state as function of $M_i$ for a
light neutrino spectrum with a NH (thick line) or IH (thin line). The dashed line at $1$ corresponds to the minimum
value for thermalisation.  }
\end{figure}

In Figure~\ref{fig:fmin} we show  the numerical results for the minimal value of $f_s(T_{\max})$ (almost identical for both species)
scanning the whole parameter space for the two sterile states, assuming their masses
differ a factor ten or more.  Varying $M_i \in [1{\rm eV}, 1 {\rm GeV}]$, we find an almost
constant value which is significantly larger than one, which means that both species 
thermalise, contributing $\Delta N_{\rm eff} \simeq 2$ when they decouple. This is the case for both neutrino hierarchies normal and inverted (NH/IH), but 
Min$[f_s(T_{max})]$ is significantly larger for IH. 
The  dependence on $M_i$ is mostly due to the
change in $g_*(T_{max})$.  

\begin{figure}
\includegraphics[scale=1]{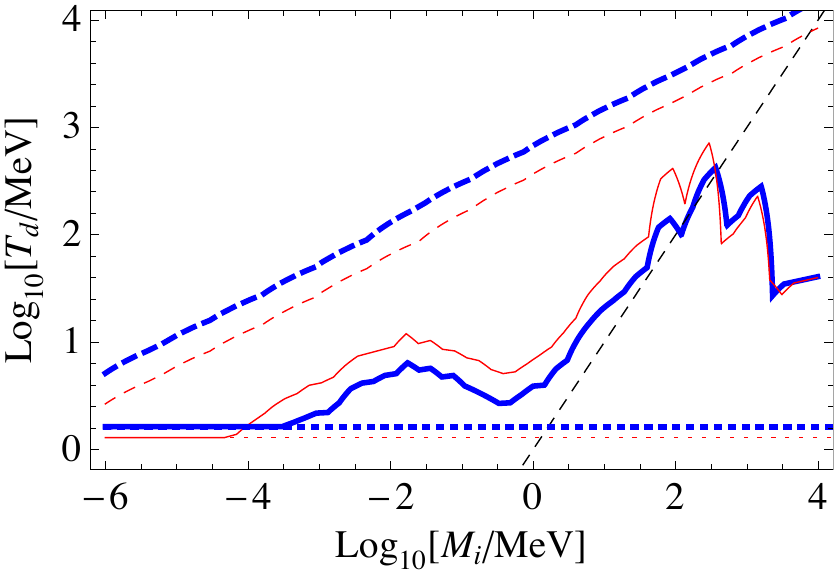}
\caption{\label{fig:T_d}  $T_d$  as function of the the sterile mass for the NH (solid
thick  line), IH (solid thin line) for parameters that minimise $f_s(T_{\rm max})$ (dashed), those that minimise $T_d$ (dotted) and those that minimise $T_d$ while 
being compatible with bounds from direct searches (solid). 
The single dashed line satisfies $T= M_i$.}
\end{figure}

We note that the thermalisation is still possible
for values of $M_i \gg 1$MeV. At some point however, the decoupling temperature 
of the sterile species will be above their mass. In this case, the contribution to $N_{\rm eff}$ requires a different treatment and will be Boltzmann
suppressed. We can estimate this   decoupling temperature, $T_{d}$, from the requirement
$f_{s}(T_d) = 1$ for $T_d < T_{\rm max}$. 
In Fig.\ref{fig:T_d} we show the value of $T_d$  as a function of $M_i$ (again the same for both species) for three cases: the  parameters that minimise 
$f_s(T_{\rm max})$ (dashed lines), the parameters that minimise $T_d$ (dotted) and the ones that minimise $T_d$ after taking into account 
direct search constraints on active-sterile mixings (solid). We see that there are regions of parameter space for all $M_i$ where sterile neutrinos remain
in equilibrium until ${\mathcal O}(1$MeV). However, as $M_i$ increases this is only possible for very special textures, inverse-seesaw like, where 
neutrino masses are suppressed due to an approximate global symmetry. Large mixings are however strongly constrained by direct searches \cite{Atre:2009rg,Ruchayskiy:2011aa}, when those bounds are included, we find that $T_d$ is well above $M_i$ for $M_i \leq {\mathcal  O}(1$GeV). If neutrinos are below this mass
they decouple when they are still relativistic, as we have assumed, and therefore contribute one unit to $\Delta N_{\rm eff}(T_d)$, but 
above this mass, they become non-relativistic before decoupling and the contribution
is suppressed by the Boltzmann factor.  

After decoupling of the sterile species, however, two important effects could modify $\Delta N_{\rm eff}$ before the active neutrino decoupling at $T_{W}$ \cite{Dolgov:2000pj}: dilution and decay. 
 
First a dilution occurs if the sterile species decouple at $T_d \gg T_{W}$, due to the change in $g_*(T)$.  The dilution can be estimated to be 
$\Delta N_{\rm eff}(T_{\rm W}) =  (g_*(T_{\rm W})/g_*(T_{d1}))^{4/3} +(g_*(T_{\rm W})/g_*(T_{d2}))^{4/3}$ provided they are still relativistic at $T_{\rm W}$ \cite{Dolgov:2000pj}. 

In order to numerically solve the kinetic equations, eq.~(\ref{eq:boltz}), we rewrite them,  as is common practice,  in terms
of the new variables \cite{Dolgov:2002wy}
\begin{align}
 x=m_0 a(t) ,\;\;\;  y=p a(t) ;
\end{align}
where $m_0$ is an arbitrary scale (fixed to be 1 MeV) and $a(t)$ is cosmic scale factor.
Equation~(\ref{eq:boltz}) becomes:
\begin{align}
\left.H(x) x \frac{\partial}{\partial x} \rho(x,y)\right|_{y}=-i[\hat{H}(x,y),\rho(x,y)]-\frac{1}{2}\{\Gamma(x,y),\rho(x,y)-\rho(x,y)_{eq}I_A\}.
\end{align}
Since we consider a range of  temperatures where  $g^*(T)$ is varying,  entropy conservation  $g^*(T(x)) T^3(x) x^3=$ {\it constant} implies that temperature does not simply scale as $\frac{1}{a(t)}$ and we take this into account. In order to avoid numerical instabilities we consider the static approximation.

\begin{figure}
\includegraphics[scale=0.8]{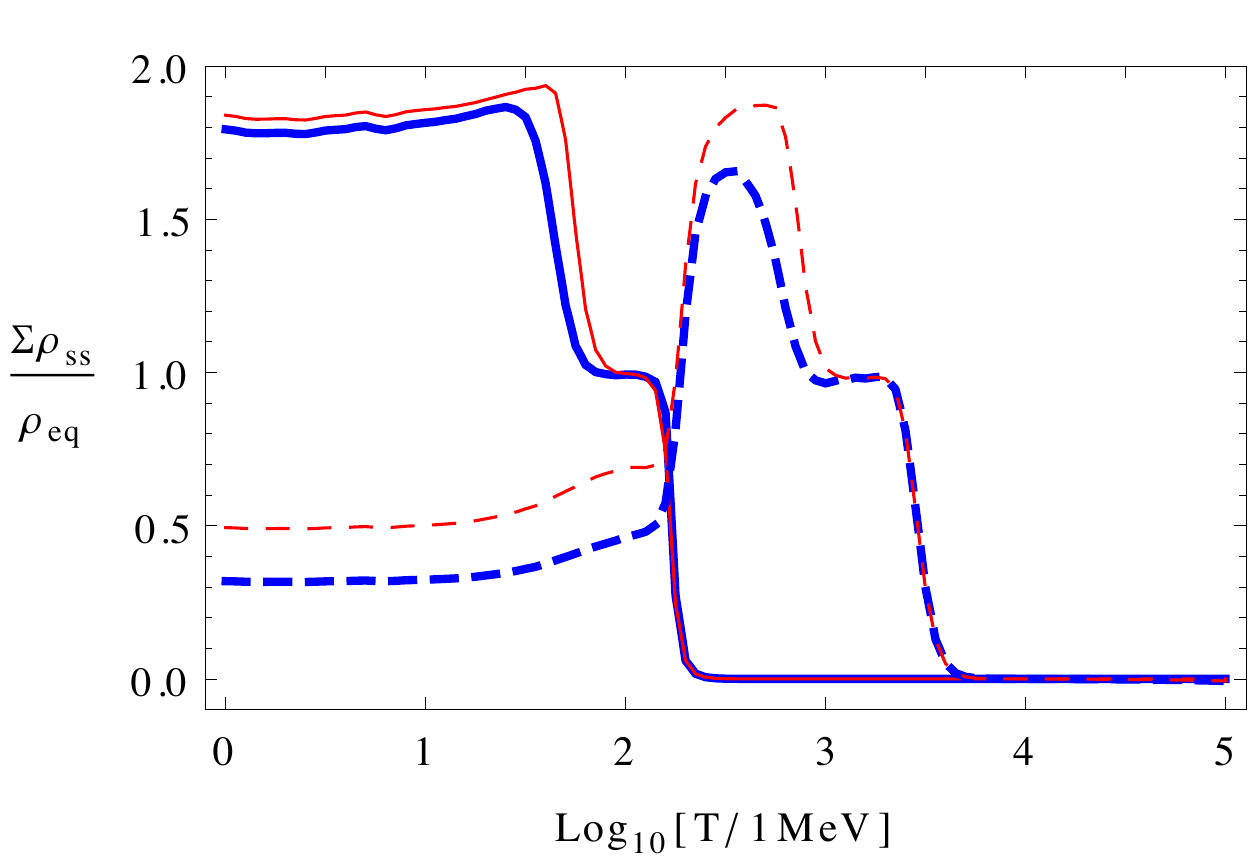}
\caption{\label{fig:exact} Evolution of the ratio of the number density of sterile species over that of  one active massless neutrino for $y=5$ for $(M_1,M_2)\simeq (2\cdot 10^{-5},10^{-3})$ (solid)  and $(0.1, 10)$ (dashed) in MeV and mixing parameters that minimize $f_{s_1}(T_{\rm max})$ for NH (thick) and IH (thin). }
\end{figure}

We have checked that, for several choices of 
mass matrix parameters, the simple estimate above gives a reasonable approximation to the numerical solution of the Boltzmann equations. The difference comes from the 
continuous change in $g^*(T)$, that we can only take into account numerically.  
In Figure~\ref{fig:exact} we show 
the evolution of the ratio of the  sterile number density  to that of one active neutrino as $T$ varies, at fixed $y=5$ and for two widely different values of $M_i$. We observe a double upward step reaching a value near  $2$ corresponding to 
the thermalisation of the two species and a dilution at lower temperatures, significant only for masses above keV.  
The dependence on $y$ of the ratio is significant due to the dilution effect and we take it into account in the definition of $\Delta N_{eff}(T_W)$ which involves the integrated energy density. 
We have considered numerically the case with degenerate heavy masses $M_1 = M_2$.  The only difference appears to be that the thermalisation curve does not show a double step but a single one. 

\begin{figure}
\includegraphics[scale=0.55]{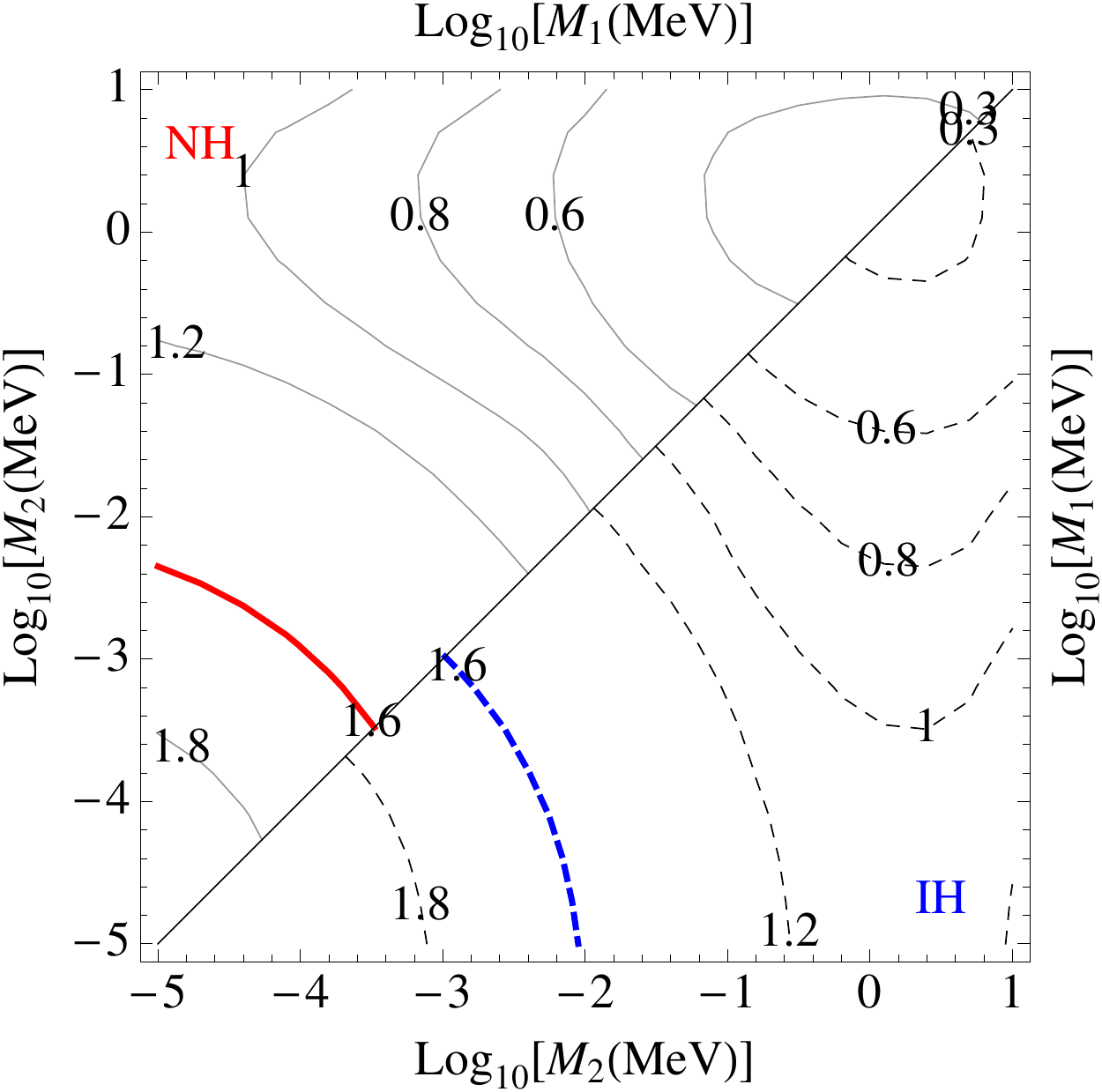}  \hspace{0.5cm} \includegraphics[scale=0.55]{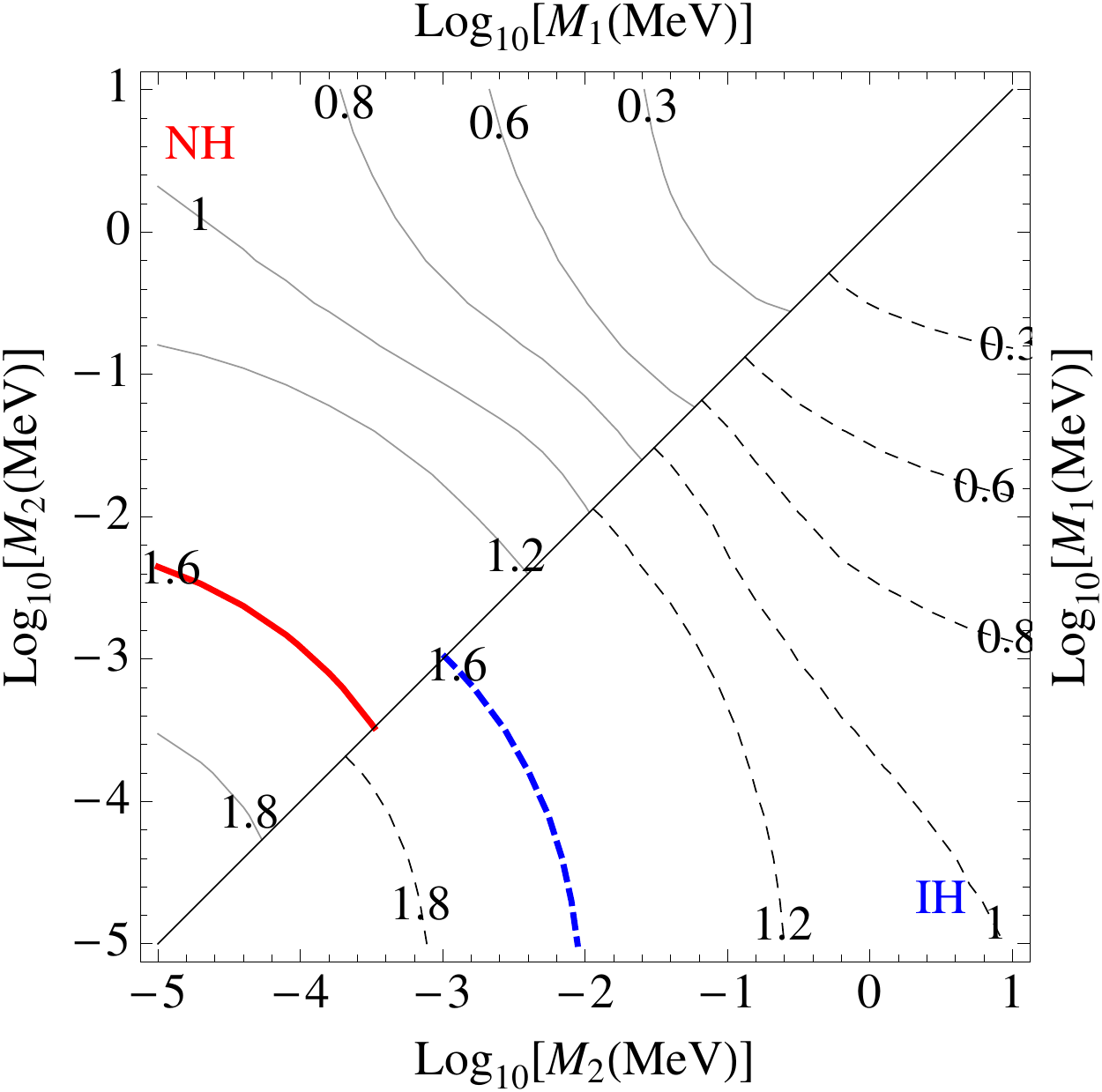}
\caption{\label{fig:dilution}  $\Delta N_{\rm eff} = \epsilon_s/\epsilon_\nu^0$ (left) and $\Delta N_{\rm eff}^{\rm rel} = (\epsilon_s- \epsilon^m_{s})/\epsilon_\nu^0$ (right) at $T_{\rm W}$ as function of the sterile masses NH  (upper octant) or IH (lower octant). The thick lines correspond to maximum allowed by BBN at 2$\sigma$.  }
\end{figure}

In Figure~\ref{fig:dilution}, we show 
the constant $\Delta N_{\rm eff}(T_{\rm W})$ lines for the mixing parameters that minimize $f_{s_1}(T_{\rm max})$, as well as those corresponding to the relativistic component, $\Delta N^{\rm rel}_{\rm eff}(T_{\rm W}) \equiv (\epsilon_s - \epsilon_s^m)/\epsilon_\nu^0$, where $\epsilon_s^m$ is the contribution of the sterile species to the matter density.   We only consider masses that remain relativistic at BBN, because more massive species would quickly dominate the energy density as cold dark matter, unless they decay before BBN.  These results show that dilution allows to relax the BBN bounds  for masses in the range 10 keV-10 MeV, however these particles give a huge contribution to the energy density when they become non relativistic at later times, modifying in a drastic way CMB and structure formation. The only way BBN and CMB  bounds could be evaded  in this range is if the sterile states decay before BBN. We come back to this point later.

We note  that the analysis might not be accurate for $T \gtrsim T_{QCD}$ \cite{Abazajian:2002yz,Asaka:2006nq}, however we do
not expect the conclusions to change drastically even if hadronic uncertainties are included.

\begin{figure}
\includegraphics[scale=0.7]{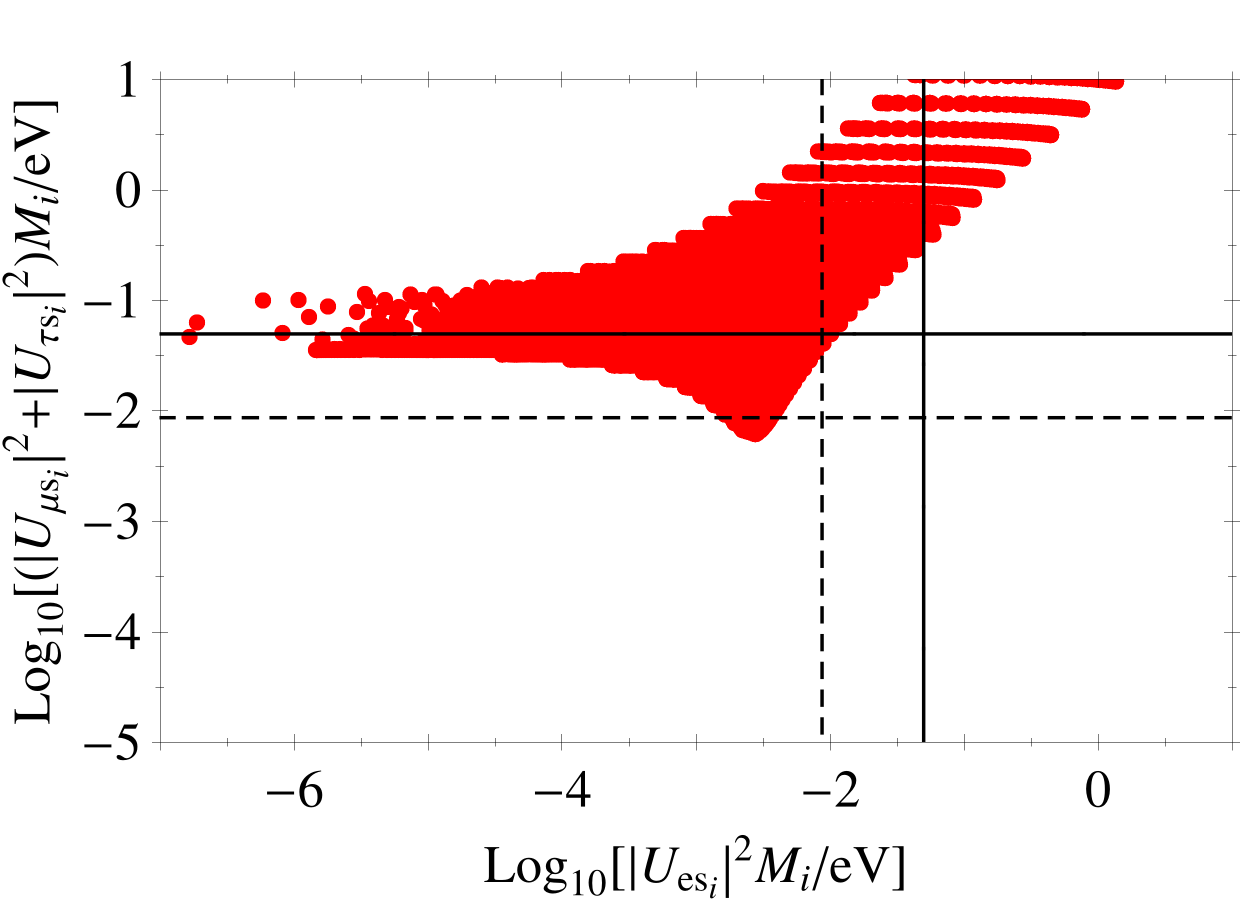}
\caption{\label{fig:unh} $(|U_{\mu s_i}|^2 + |U_{\tau s_i}|^2) M_i$ versus $|U_{e s_i}|^2 M_i$ varying the unconstrained paremeters for NH.  The solid/dashed
line  corresponds to $m_3 \sim \sqrt{\Delta m^2_{\rm atm}}$/$m_2\sim \sqrt{\Delta m^2_{\rm solar}}$.}
\label{fig:cornh}
\end{figure}

It is important to stress that the approximate independence of thermalisation on the heavy masses $M_i$ results from the approximate
seesaw scaling of the  $|U_{a s_i}|^2 M_i \sim m_l$, which is only approximate since there is dependence on several 
unknown parameters, see eq.~(\ref{eq:param}).   Fig.~\ref{fig:cornh} shows the values of $|U_{e s_i}|^2 M_i$ and
$(|U_{\mu s_i}|^2 + |U_{\tau s_i}|^2) M_i$ within the full range of the unconstrained parameters for the normal  
hierarchy. 
We note that  $|U_{e s_i}|^2 M_i$
 can get  extremely small. Had we only considered the oscillations to electrons in this case, we would have found that for those
 parameters $f_s(T_{\rm max})\ll 1$, but
$(|U_{\mu s_i }|^2 + |U_{\tau s_i}|^2) M_i$  is in the expected ballpark and therefore the thermalisation takes place through 
the oscillation to $\mu$ and $\tau$. A similar pattern is observed for the IH, both combinations do not 
get very small simultaneously. 

For sufficiently high mass the sterile neutrino could decay before BBN and our analysis is not valid for this situation.  The lifetime is in the range $\tau  \sim 6 \times 10^{11}  \left[ {{\rm MeV}\over M_i}\right]^4 \left[ \frac{ 0.05  eV}{ |U_{as_i}|^2 M_i}\right] s$, below the $\pi_0$ threshold, which means they decay  after BBN below this threshold, for natural choices of mixings. However, the mixings might reach values
significantly larger  (see Figure~\ref{fig:cornh}).  For extreme mixings of $O(1)$, neutrinos as light as 10~MeV  could decay before BBN. The bounds on short-lived  sterile neutrinos with masses in the range [10 MeV,140 MeV]  have been studied in \cite{Dolgov:2000pj,Fuller:2011qy,Ruchayskiy:2012si} and very strong bounds have been found combining BBN and direct accelerator searches, essentially excluding this possibility  \cite{Ruchayskiy:2011aa}.  The analysis above 140 MeV  gets more complicated with various competing effects that occur near the QCD phase transition. 

We want to stress however that  in the generic seesaw models that we are considering, such short lifetimes  result only from very specific textures in which an approximate global symmetry (and not small Yukawa couplings) suppresses light neutrino masses in front of the seesaw scale. 
The flavour structure of these models is even more constrained, but large active-sterile mixings can be reached. Note that in these corners
of parameter space, thermalisation will be more efficient and $T_d$ will be closer to $T_W$, so dilution is less relevant.  

The previous results show that  the sterile states in generic low-scale seesaw $3+2$ models do thermalize independently of  the scale of the Majorana masses, within a wide
range. This implies very strong constraints from cosmology. The following conclusions can be drawn.

1) $M_{1,2} \lesssim {\mathcal O}(100 {\rm MeV})$:   $\Delta N_{\rm eff}(T_d) \simeq 2$ and decay after BBN, which is incompatible with the present BBN or/and CMB constraints independently of the mass 
of the sterile states. These models are therefore strongly disfavoured. 

2) $M_1 \lesssim  {\mathcal O}(100 {\rm MeV})$, $M_2\gtrsim {\mathcal O}({\rm GeV})$: $\Delta N_{\rm eff}(T_d) \simeq 1$, while the heavy state is Boltzmann suppressed at decoupling or decays before $T_{\rm W}$. BBN constraints can accommodate this case if $M_1$ is still relativistic at BBN. However CMB and LSS measurements  close this window all the way down to $M_1 \leq 0.36~ {\rm eV}$ or so at $95\%$CL \cite{DiValentino:2013qma}.

3) $M_{1,2} \gtrsim {\mathcal O}(1 {\rm GeV})$ survive at present cosmological constraints on $N_{\rm eff}$, because they decouple while being non-relativistic and therefore $\Delta N_{\rm eff}(T_d)$ is Boltzmann suppressed, or because they decay well before $T_{\rm W}$. 

Establishing precisely what happens  in the range 100~MeV-1~GeV, specially in the case of approximate global symmetries (large mixings) where neutrinos could decay before BBN requires a more complex analysis. A strong dependence on the unknown mixing parameters is to be expected in this range. 

\begin{acknowledgments}

We thank A.Donini, J. Lesgourgues, O. Mena, C. Pe\~na-Garay, J. Racker, N. Rius and J. Salvado for useful discussions. We warmly thank E. Fern\'andez-Martinez
for pointing out an error in an earlier version of the paper. 
This work was partially supported by grants FPA2011-29678, PROMETEO/2009/116, CUP (CSD2008-00037) and  ITN INVISIBLES (Marie Curie Actions, PITN-GA-2011-289442). 

\end{acknowledgments}


\vspace{0.5cm}
\begin{thebibliography}{99}

\bibitem{Kopp:2013vaa} For a recent review see
  J.~Kopp {\it et al},    JHEP {\bf 1305} (2013) 050.
    
    \bibitem{Aguilar:2001ty}
  A.~Aguilar-Arevalo {\it et al.}  [LSND Collaboration],
  Phys.\ Rev.\ D {\bf 64} (2001) 112007.
    
    \bibitem{Aguilar-Arevalo:2013pmq}
  A.~A.~Aguilar-Arevalo {\it et al.}  [MiniBooNE Collaboration],
  Phys.\ Rev.\ Lett.\  {\bf 110} (2013) 161801.
  
  \bibitem{reactor}
  G.~Mention {\it et al}, 
  Phys.\ Rev.\ D {\bf 83} (2011) 073006.
  P.~Huber,
  Phys.\ Rev.\ C {\bf 84} (2011) 024617
   [Erratum-ibid.\ C {\bf 85} (2012) 029901].
  
   \bibitem{Dodelson:1993je}
  S.~Dodelson and L.~M.~Widrow,
  Phys.\ Rev.\ Lett.\  {\bf 72} (1994) 17.

\bibitem{Akhmedov:1998qx}
  E.~K.~Akhmedov, V.~A.~Rubakov and A.~Y.~Smirnov,
  Phys.\ Rev.\ Lett.\  {\bf 81} (1998) 1359.

  \bibitem{miniseesaw}
  A.~de Gouvea,
  Phys.\ Rev.\ D {\bf 72} (2005) 033005.
  A.~de Gouvea, J.~Jenkins and N.~Vasudevan,
  Phys.\ Rev.\ D {\bf 75} (2007) 013003.
  
    \bibitem{Donini:2011jh}
  A.~Donini {\it et al},
  JHEP {\bf 1107} (2011) 105.
  
  \bibitem{Donini:2012tt}
  A.~Donini {\it et al},
  JHEP {\bf 1207} (2012) 161.

  
  
   
\bibitem{Canetti:2012kh}
  L.~Canetti {\it et al},
  Phys.\ Rev.\ D {\bf 87} (2013) 093006 and references therein.
    
\bibitem{Mangano:2005cc}
  G.~Mangano {\it et al},
  Nucl.\ Phys.\ B {\bf 729} (2005) 221.
    
\bibitem{Izotov:2010ca}
  Y.~I.~Izotov and T.~X.~Thuan,
  Astrophys.\ J.\  {\bf 710} (2010) L67.
  
  \bibitem{Ade:2013zuv}
  P.~A.~R.~Ade {\it et al.}  [Planck Collaboration],
  arXiv:1303.5076 [astro-ph.CO].
 
 \bibitem{WMAP}
  G.~Hinshaw {\it et al.}  [WMAP Collaboration],
  Astrophys.\ J.\ Suppl.\  {\bf 208} (2013) 19.
  
  
\bibitem{SPT}
  Z.~Hou {\it et al.},
  arXiv:1212.6267 [astro-ph.CO].




\bibitem{ACT}  
  J.~L.~Sievers {\it et al}, 
  arXiv:1301.0824 [astro-ph.CO].
  
 
  
    \bibitem{Dolgov:2003sg}
  A.~D.~Dolgov and F.~L.~Villante,
    Nucl.\ Phys.\ B {\bf 679} (2004) 261.
  
  \bibitem{Cirelli:2004cz}
  M.~Cirelli, G.~Marandella, A.~Strumia and F.~Vissani,
    Nucl.\ Phys.\ B {\bf 708} (2005) 215
  [hep-ph/0403158].
  
    \bibitem{Melchiorri:2008gq}
  A.~Melchiorri {\it et al},
    JCAP {\bf 0901} (2009) 036.

 \bibitem{Hannestad:2012ky}
  S.~Hannestad, I.~Tamborra and T.~Tram,
    JCAP {\bf 1207} (2012) 025.

  \bibitem{Kuflik:2012sw}
  E.~Kuflik, S.~D.~McDermott and K.~M.~Zurek,
    Phys.\ Rev.\ D {\bf 86} (2012) 033015.


  \bibitem{Jacques:2013xr}
  T.~D.~Jacques, L.~M.~Krauss and C.~Lunardini,
    Phys.\ Rev.\ D {\bf 87} (2013) 083515.
  
  \bibitem{Archidiacono:2013xxa}
  M.~Archidiacono {\it et al}
   arXiv:1302.6720 [astro-ph.CO].

 \bibitem{Casas:2001sr}
  J.~A.~Casas and A.~Ibarra,
  Nucl.\ Phys.\ B {\bf 618} (2001) 171.

 \bibitem{Blennow:2011vn}
  M.~Blennow and E.~Fernandez-Martinez,
  Phys.\ Lett.\ B {\bf 704} (2011) 223.
  
   \bibitem{simple:1990} R.~Barbieri and A.~Dolgov,
  Phys.\ Lett.\ B {\bf 237} (1990) 440.   K.~Kainulainen,
  Phys.\ Lett.\ B {\bf 244} (1990) 191.
  
  \bibitem{Notzold:1987ik}
  D.~Notzold and G.~Raffelt,
  Nucl.\ Phys.\ B {\bf 307} (1988) 924. 
  
  \bibitem{ys}
  A.~D.~Dolgov, S.~H.~Hansen, S.~Pastor and D.~V.~Semikoz,
  Astropart.\ Phys.\  {\bf 14} (2000) 79.

  
   \bibitem{Sigl:1992fn}
  G.~Sigl and G.~Raffelt,
  Nucl.\ Phys.\ B {\bf 406} (1993) 423.

\bibitem{Dolgov:2002wy}
  A.~D.~Dolgov,
  Phys.\ Rept.\  {\bf 370} (2002) 333
  [hep-ph/0202122].
   
    
    \bibitem{KTY}
  K.~Kimura, A.~Takamura and H.~Yokomakura,
  Phys.\ Lett.\ B {\bf 537} (2002) 86 and 
  Phys.\ Rev.\ D {\bf 66} (2002) 073005.
   A.~Donini {\it et al}, 
  JHEP {\bf 0908} (2009) 041.
  
    \bibitem{Abazajian:2002yz}
  K.~N.~Abazajian and G.~M.~Fuller,
    Phys.\ Rev.\ D {\bf 66} (2002) 023526. 
  K.~Abazajian,
  Phys.\ Rev.\ D {\bf 73} (2006) 063506. 

\bibitem{Asaka:2006nq}
  T.~Asaka, M.~Laine and M.~Shaposhnikov,
  JHEP {\bf 0701} (2007) 091.

\bibitem{Dolgov:2000pj}
  A.~D.~Dolgov, S.~H.~Hansen, G.~Raffelt and D.~V.~Semikoz,
  Nucl.\ Phys.\ B {\bf 580} (2000) 331 and   Nucl.\ Phys.\ B {\bf 590} (2000) 562.

     \bibitem{Fuller:2011qy}
  G.~M.~Fuller, C.~T.~Kishimoto and A.~Kusenko,
  arXiv:1110.6479 [astro-ph.CO].

   \bibitem{Ruchayskiy:2012si}
  O.~Ruchayskiy and A.~Ivashko,
  JCAP {\bf 1210} (2012) 014.


\bibitem{Atre:2009rg}
  A.~Atre, T.~Han, S.~Pascoli and B.~Zhang,
  JHEP {\bf 0905} (2009) 030.

\bibitem{Ruchayskiy:2011aa}
  O.~Ruchayskiy and A.~Ivashko,
  JHEP {\bf 1206} (2012) 100.

    
  
  \bibitem{DiValentino:2013qma}
  E.~Di Valentino, A.~Melchiorri and O.~Mena,
  arXiv:1304.5981 [astro-ph.CO].
  


\end{thebibliography}
\end{document}